# Deterministic separation of suspended particles in a reconfigurable obstacle array

*Siqi Du and German Drazer*

*Mechanical and Aerospace Engineering Department, Rutgers, The State University of New Jersey, Piscataway, NJ.*

## Abstract

We use a macromodel of a flow-driven deterministic lateral displacement (DLD) microfluidic system to investigate conditions leading to size-separation of suspended particles. This model system can be easily reconfigured to establish an arbitrary orientation between the average flow field and the array of obstacles comprising the stationary phase (forcing angle). We also investigate the effect of obstacle size using two arrays with different obstacles but same surface-to-surface distance between them. In all cases, we observe the presence of a locked mode at small forcing angles, in which particles move along a principal direction in the lattice until a locked-to-zigzag transition takes place when the driving force reaches a critical angle. We show that the transition occurs at increasing angles for larger particles, thus enabling particle separation at specific forcing angles. Moreover, we observe a linear correlation between the critical angle and the size of the particles that could be used in the design of microfluidic systems with a fixed orientation of the flow field. Finally, we present a simple model, based on the presence of irreversible interactions between the suspended particles and the obstacles, which describes the observed dependence of the migration angle on the orientation of the average flow.

## Introduction

Microfluidic methods that fractionate mixtures into individual chemical or biological components constitute an integral part in micro-total-analysis-systems (μ-TAS). These methods can be broadly classified as *active* or *passive* depending on the use or not of an external field to drive the separation. Active methods include dielectrophoresis [1], magnetophoresis [2], acoustophoresis [3], various optical methods [4], [5] and a family of field flow fractionation methods with different fields driving the separative displacement [6]–[9]. Passive methods are generally based on hydrodynamics and particle-solid interactions between the species and the stationary phase in the fluidic system [10], [11], including hydrodynamic filtration [12], pinched flow fraction [13]–[16] and separation by inertial and Dean flows [17]. Deterministic lateral displacement (DLD) is a separation method that can be implemented in both active and passive modes. Introduced as a passive method [18], DLD exploits the experimental observation that particles of different sizes flowing through a periodic array of cylindrical obstacles may migrate in different directions, thus leading to separation (Figure 1). In addition to size-based separation, recent work suggests that DLD could also effectively fractionate a mixture depending on shape [19], [20]. The ability to continuously separate particles of different size or shape without the need for an external force field has made DLD suitable for the fraction of a number of biological samples [21]–[27]. Although DLD was initially introduced as a flow-driven, passive microfluidic method for size separation, we have shown in previous work that driving the particles by external forces also results in separation depending on the orientation of the force with respect to the array of obstacles. Specifically, we have successfully used gravity and electric fields to drive the separation of suspended particles in force-driven DLD (f-DLD) [28]–[31]. In their original work, Inglis *et al.* observed two different types of trajectories, depending on the size of the particles: *bump mode* trajectories for larger particles and *zigzag mode* trajectories for smaller ones [32]. The former, represented by the bigger circles in Figure 1, corresponds to a situation where the particle moves alongside a line of obstacles on the lattice (a *column of obstacles*), and the latter, depicted by the smaller circles in Figure 1, corresponds to trajectories in which the particle zigzags inside the array of. Inglis *et al.* postulated that *zigzag mode* trajectories were, on average, aligned with the flow direction. However, Kulrattanarak *et al.* later reported that when particles zigzag inside the array, in general, do not necessarily move in the direction of the flow [33], [34]. In all cases, as the size of the particles increases, their motion eventually transitions from *zigzag mode* to *bump mode* for particles larger than a critical size, a transition that depends on the geometry of the lattice. In previous work, we have shown that f-DLD results in similar behavior [35]. Moreover, using scaled-up macromodels of microfluidic DLD systems, we were able to investigate the entire range of possible orientations of the external force with respect to the array of obstacles, and show that, in fact, all particles transition from *bump mode* to *zigzag mode* as the forcing angle increases [28], [35], [36]. In order to better describe the properties of the observed trajectories, let us first define some characteristic angles. First, we define the forcing angle α as the angle between a column of obstacles in the lattice and the external force (or average flow). We then define the migration angle β as the angle between the average migration of the particles and a column in the array (see Figure 1). L denotes the direction of particle migration, F represents the direction of the flow and C gives the direction of the lattice. The dashed line L' that is parallel to L is drawn to better illustrate the migration angle β. The macromodel experiments showed that, as the forcing angle increases from zero, all the particles remain *locked* to move alongside a column of obstacles in the array, i.e. β=0°, until they reach a critical forcing angle, $\alpha_c$, after which they are able to move across columns of obstacles, resulting in a periodic zigzag motion, with α ≠β in general. In fact, for any forcing angle, the motion of the particles is periodic and the average migration is always in certain lattice directions. Then considering the existence of non-hydrodynamic interactions between suspended particles and solid obstacles, we developed a simple geometric model that captured the observed dynamics [28], [35], [37].

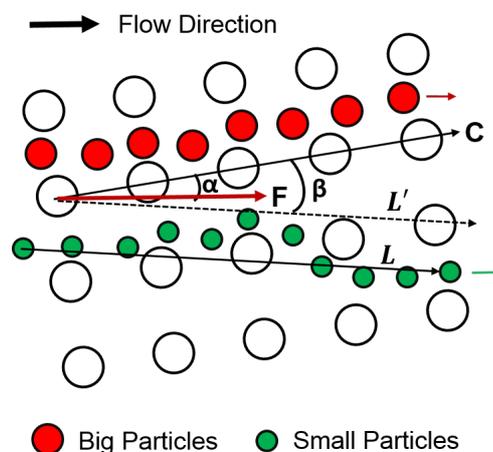

Figure 1 Schematic view of DLD. The circles represent the positions of one particle at different time of movements. The solid line L denotes the direction of particle migration, the arrow F represents the direction of the flow and the solid line that connects centers of a column of obstacles gives the direction of the lattice. The dashed line L' that is parallel to L is drawn to better illustrate the migration angle β.

Here, we use macromodels of flow-driven DLD devices to investigate conditions leading to size-separation of suspended particles depending on the geometry of the array of obstacles and the average orientation of the pressure-driven flow. In previous experiments using classical flow-driver DLD microfluidic devices, the forcing angle is built into the system and cannot be modified. As a result, such devices were only able to fractionate a sample between particles larger and smaller than a critical size that is fixed by the geometry of the system. In the present study, we continuously vary the orientation angle and, therefore, interrogate particles of different size over the entire range of driving forces. We also use two different arrays, differing in the size of the obstacles. Based on this exhaustive set of experiments, we shall show that, the motion of suspended particles is analogous that observed in the f-DLD case. Specifically, we experimentally show: (i) the existence of a *locked mode* for all particle sizes, in which the average migration angle is β=0° and particles move along a column of obstacles; (ii) a sharp transition from *locked mode* to *zigzag mode* in which particles move periodically at certain lattice directions; (iii) a monotonic increase in the critical angle at which the *locked-to-zigzag* transition occurs with particle size. Finally we present a simple model based on the irreversible nature of non-hydrodynamic interactions between the suspended particles and the obstacles that accordingly describes the observed dependence of the migration angle on the orientation of the average flow.

## Experiment Set-up

Our experimental set-up is a scaled-up version of a microfluidic DLD system, consisting of an acrylic rectangular channel of width L = 280mm (see Figure 2A). A square array of obstacles is centered in the channel and, the central part of the array can be oriented at any angle with respect to the direction of the average flow along the channel, as shown in Figure 2B. The moving central part allows us to vary the forcing angle continuously over the entire range of possible orientations with the same system. Additionally, to study the effect of obstacle size on particle trajectory, two different arrays are used. As showed in Figure 2C, the difference between the two arrays is the size of the obstacles, either 1mm or 2mm in diameter. The height of the channel (and obstacles) is 5mm and the open gap between obstacles is 4mm, in both arrays. Both arrays were fabricated using a 3D printer (Objet350 Connex, Stratasys). The reason for the rectangular shape of the complete array is to ensure a uniform flow over the width of the channel (except close to the walls). A circular array alone, in contrast, would not provide a uniform flow resistance over the width of the channel and would lead to large flow variations. The flow is driven by a constant pressure drop generated by a Mariotte's bottle and distributed over the channel width at the inlet using a manifold.

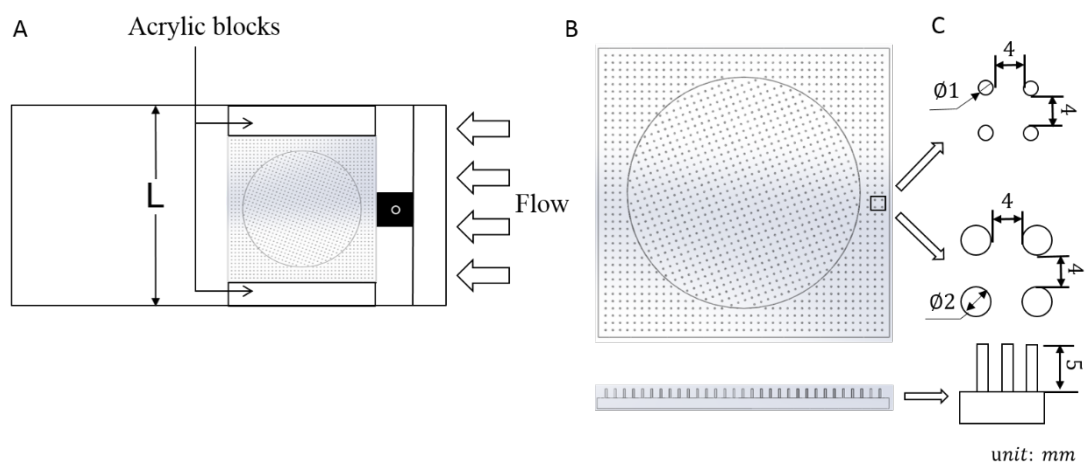

Figure 2 A. Schematic view of the experiment setup. B. Top and side view of the array of obstacles. C. We use two different arrays with different obstacles diameter as indicated. The height of the obstacles and the surface to surface gap between two obstacles are the same in both arrays.

In order to compare our results with microfluidic systems we have to satisfy both geometric and dynamic similarities. Therefore, in order to maintain low Reynolds numbers, comparable to those common in microfluidics, we use a mixture of glycerin (99% Glycerin, McMaster-Carr) and water with a volume ratio 3:2. The kinematic viscosity of the liquid mixture is approximately $1.38 \times 10^{-5}$ m$^2$/s [38]. The flow rate in each experiment is approximately 8 cm$^3$/s, and the corresponding Reynolds number can be estimated as $Re = vh/v \approx 4$ , where $h=5$ mm is the height of the channel. We use Nylon particles of six different sizes, with diameters 1/16" (1.59 mm), 3/32" (2.38 mm), 1/8" (3.18 mm) (McMaster-Carr), 0.072" (1.83 mm), 7/64" (2.78 mm) and 9/64" (3.57 mm) (Precision Plastic Ball Co.).

Each experiment consists of a given particle size and a given forcing angle. We analyze the trajectories of several particles, between 20 and 30 depending on variability, and determine the average migration angle. We then repeat the procedure for each particle size in the two different arrays and for several forcing angles.

## Results and Discussion

First, we investigate the presence of a locking mode, in which particles move along a column of obstacles (β=0°) for forcing angles lower than a certain critical angle and, as the forcing angle increases beyond the critical value, the motion of the particles sharply transitions into *zigzag mode*. To analyze the presence of locking and the *locked-to-zigzag mode* transition we introduce the probability of crossing $P_c$. For a given forcing angle we define the probability of crossing as the fraction of particles that move in *zigzag mode* over the total number of particles analyzed. Alternatively, $1 - P_c$ is the fraction of particles locked to move in the [1, 0] lattice direction with β=0° and, therefore, do not move across columns of obstacles. The results are presented in Figure 3 for the two different obstacle arrays and all particle sizes. Clearly, in all cases, we observe a sharp transition from no crossing (i.e. locked mode at β=0°) to complete crossing with $P_c = 1$. Therefore, we define the critical angle for each particle as the angle at which the crossing probability is $P_c = 1/2$.

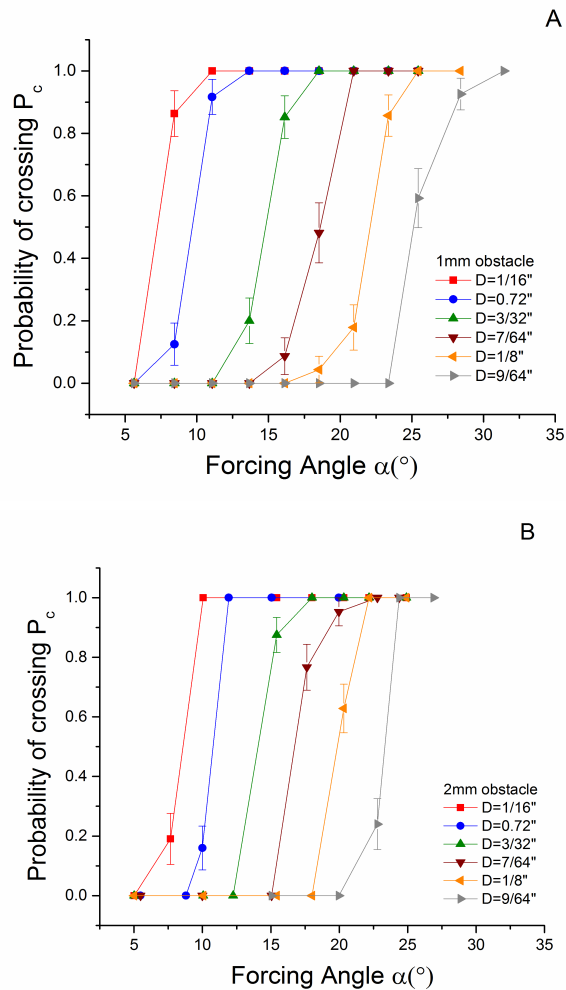

Figure 3 A. Probability of crossing for different size of particles for 1 mm obstacle array. B. Probability of crossing for different size of particles for 2 mm obstacle array.

It is also clear in Figure 3 that the *locked-to-zigzag* transition occurs at increasing forcing angles for particles of increasing size. This demonstrates that particles can be separated by size, based on differences in their critical angle, and that excellent resolution is ensured by the sharp transitions into *zigzag mode*. In Figure 4 we present the critical angle as a function of particle size for the two different arrays of obstacles. Interestingly, we observe a linear relationship extending the entire range of forcing angles. This linear trend is in contrast with the non-linear dependence predicted by Inglis and coworkers in their model [32], represented by the solid line in the plot.

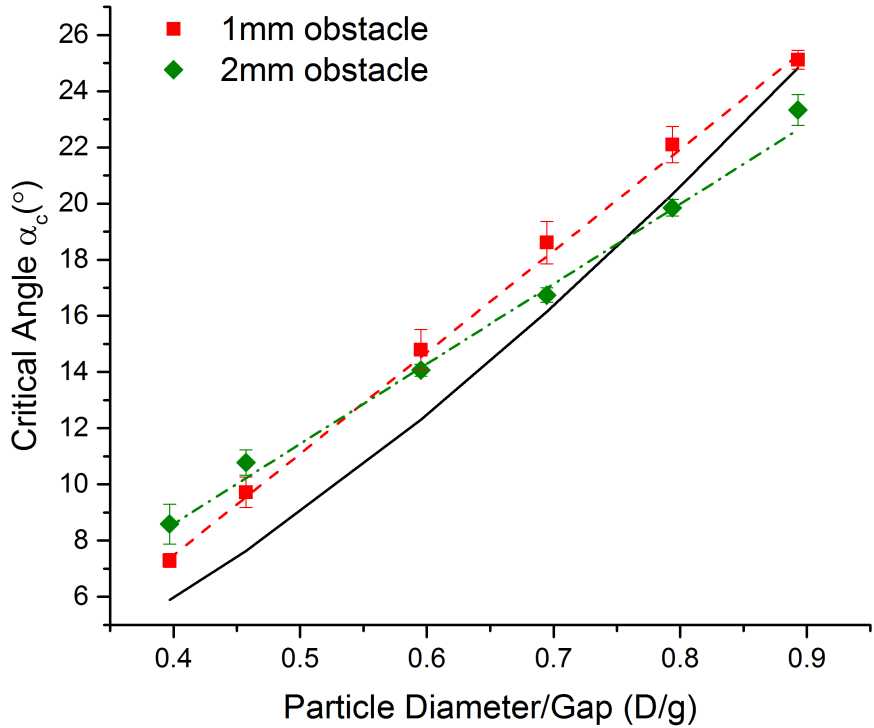

Figure 4 Critical angles for different size of particles in both arrays. Particle diameter is normalized by the gap size. Straight lines correspond to a linear fit of the results. The solid curve is calculated using Inglis's model. For 1mm obstacle array, the fitting result is $\alpha_c = 36.2861(D/g) - 6.90216, R^2 = 0.9986$, and for 2mm obstacle array the fitting result is $\alpha_c = 28.7769(D/g) - 2.81352, R^2 = 0.9956$.

Second, we investigate the migration angle in the *zigzag mode*. As we discussed in the introduction, the original DLD work assumed that particles move, on average, parallel to the forcing angle. However, further analysis showed that this might not be the case [33], [34]. Similarly, our f-DLD experiments clearly indicate that particles move periodically at specific lattice directions that, in general, are not aligned with the external force [28], [35]. In Figure 5, we present the migration angle as a function of the forcing angle. We clearly observe the *locked-to-zigzag* transition and, in most cases, the migration angles in the *zigzag mode* are similar to the forcing angle ($\beta = \alpha$, as indicated by the dotted line in the plots). However, some differences between the migration and forcing angles can be observed, as suggested by previous simulation and experimental work [33], [34], [39]. We will discuss these deviations in the next section.

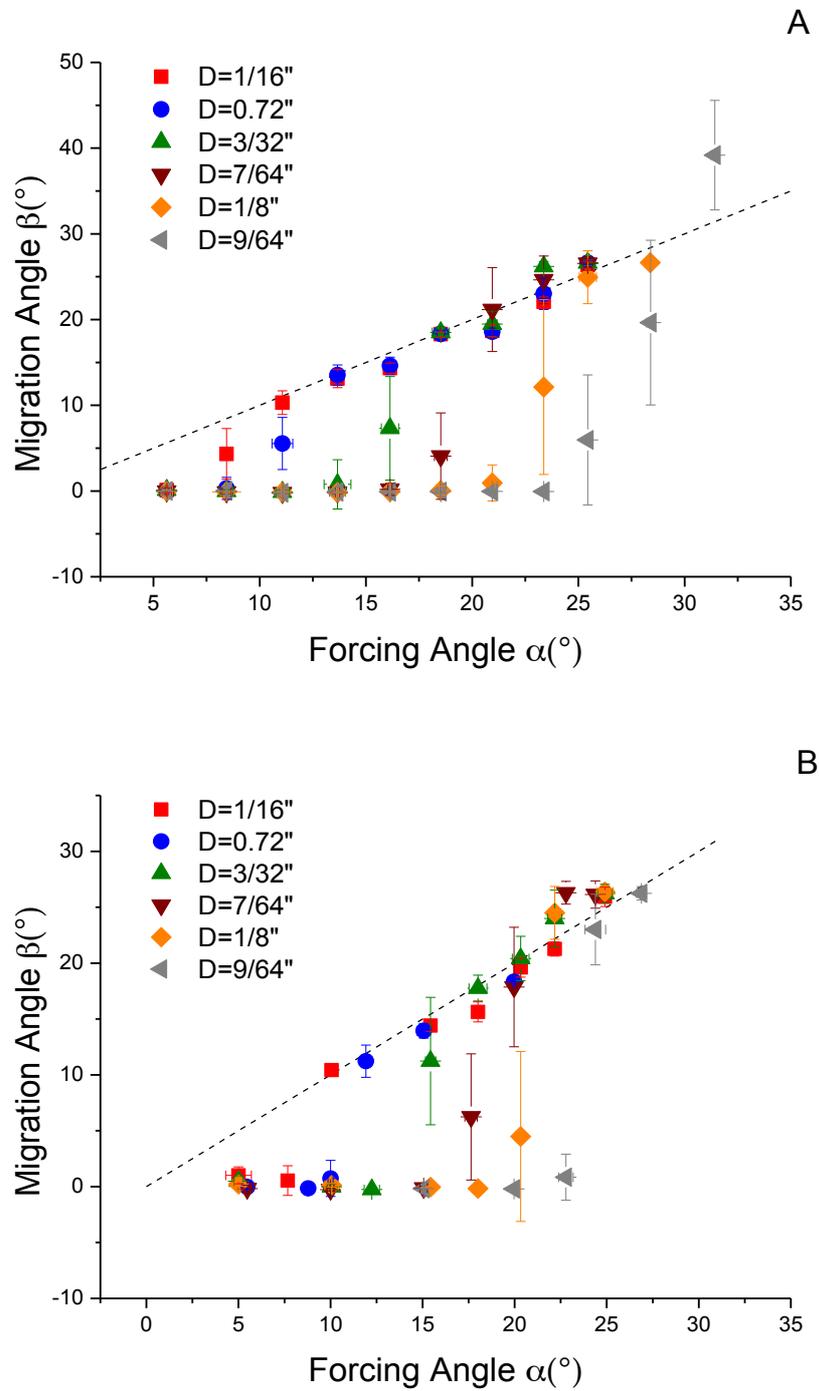

Figure 5 A. Migration angle as a function of forcing angle in 1 mm obstacle array. B. Migration angle as a function of forcing angle in 2 mm obstacle array. The dashed line in A and B is $\beta = \alpha$.

## Migration Model

We propose a model based on the individual particle-obstacle interactions as the suspension moves through the array of obstacles. We refer to the motion of a suspended particle around and past an obstacle as a particle-obstacle *collision*. Moreover, we assume that during such a collision and depending on the initial offset, particles experience irreversible interactions that lead to a net lateral displacement. (Note that these assumptions (and the resulting model) are completely analogous to those used to describe f-DLD systems, and a more detailed discussion can be found elsewhere [35], [37], [40].) In fact, collisions can be divided into two groups depending on the initial offset, as schematically shown in Figure 6. Collision for which the initial offset $b_{in}$ is larger than a certain critical value $b_c$ ($b_{in} > b_c$), are *reversible* and there is no net lateral displacement resulting from the collision. In this case, trajectories are fore-and-aft symmetric. On the other hand, collisions for which the initial offset is small (see shaded region in Figure 6) ($b_{in} < b_c$), are *irreversible* and their outgoing offset is $b_c$. In other words, irreversible collisions result in a net lateral displacement of magnitude ($b_c - b_{in}$). Note that the resulting collapse of irreversible trajectories, in which any incoming particle with $b_{in} < b_c$, that is inside the shaded region in Figure 6, comes out of the collision with the same offset $b_c$, implies the existence of *directional locking*, with certain migration angles that remain constant for a finite range of forcing angles [41], [36], [37].

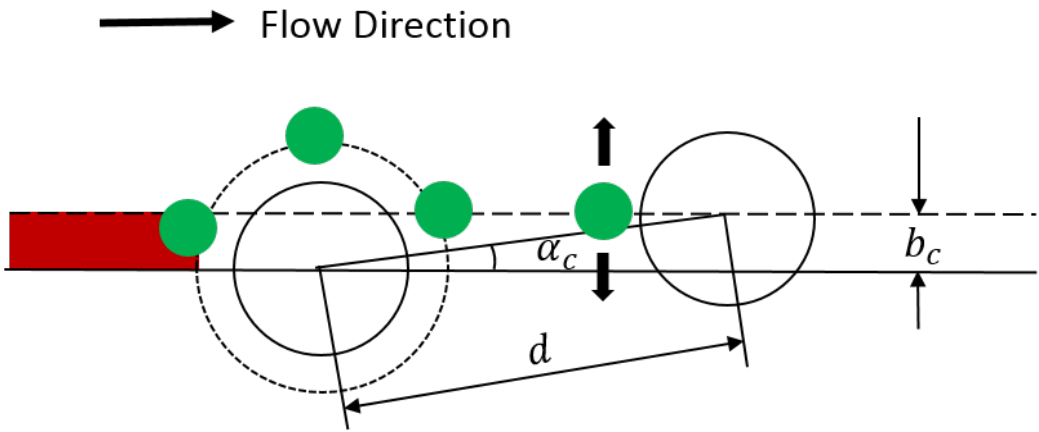

Figure 6 Schematic view of a particle obstacle collision showing the case in which the forcing angle is exactly at the critical value. The dashed line depicts the height of the center of a particle coming out of the collision with respect to the centerline of the obstacle, which is exactly the critical offset for this particular particle.

The critical offset is the only parameter in the proposed collision model, for a given particle size and geometry of the obstacle array, and can be obtained from the critical angle determined by the crossing probability using geometry relation $b_c = d \, sin(\alpha_c)$. After the critical offset is determined, it is straightforward to calculate the migration angle as a function of the forcing angle from geometric considerations, given that the result of every particle-obstacle collision can be predicted. In fact, only those collisions that are irreversible need to be accounted for and they simply result in a net lateral displacement perpendicular to the forcing direction [37], [41]. In Figure 7, we show the comparison between the proposed model and the experimental results. Note that, although the critical offset is determined from the crossing probability, these results are not independent from the migration angle experiments and the comparison presented in Figure 7 should be interpreted as a partial fit to the data. That is, although the critical angle is determined by fitting the data, the comparison for forcing angles above the critical angle has no fitting parameter. In general, we observe good agreement between the model and the experimental results, which suggests that in the *zigzag mode* the migration angle is not necessarily the same as the forcing angle but rather corresponds to directional locking into certain lattice directions.

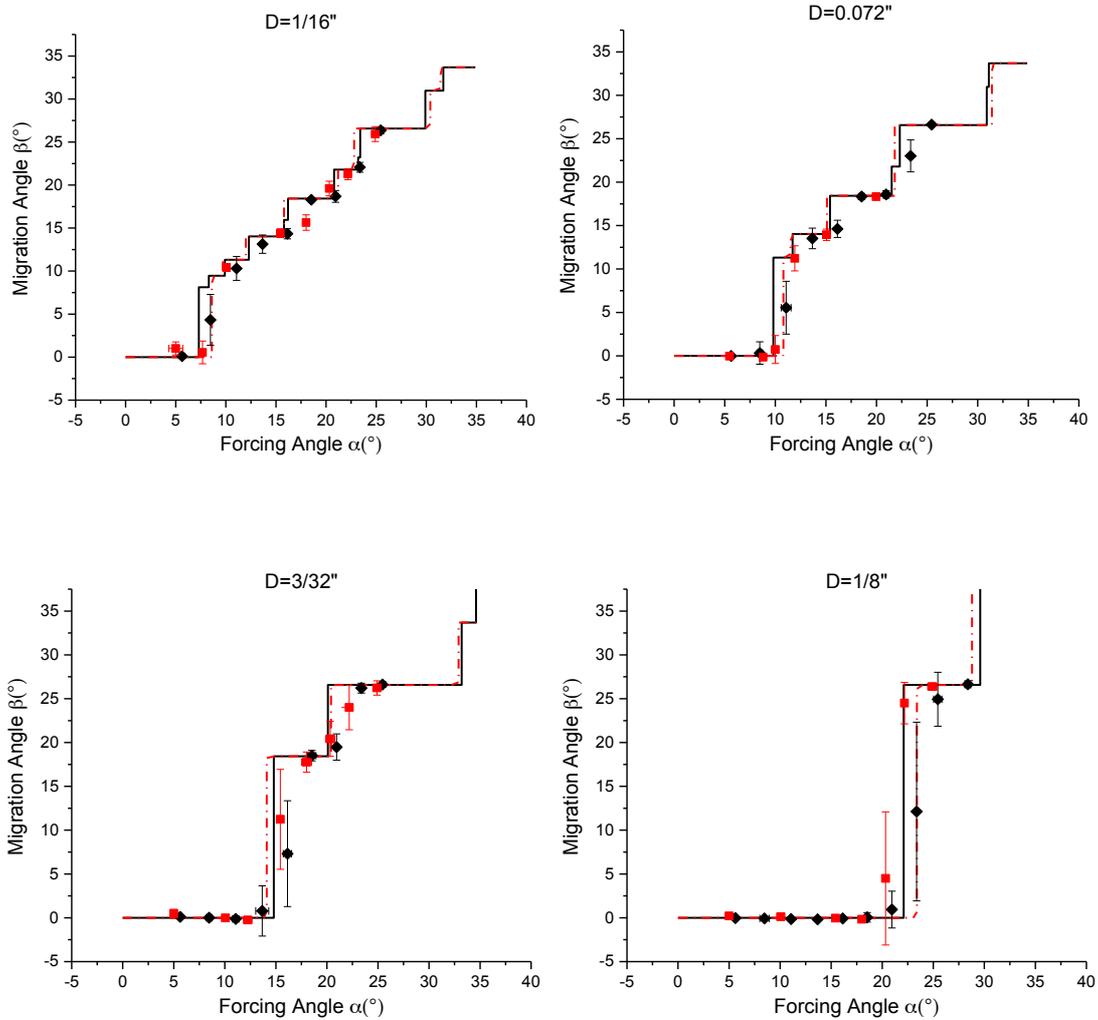

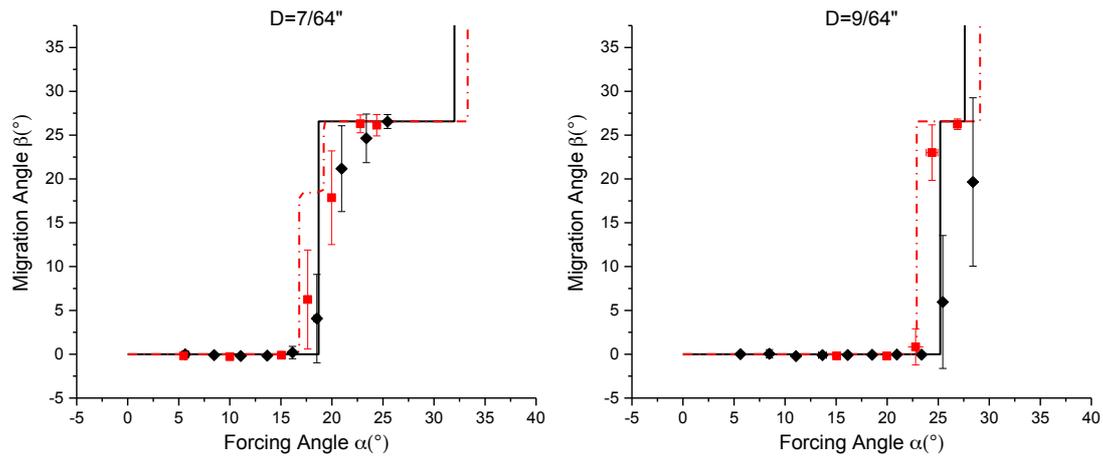

Figure 7 Model fitting results for particles of all sizes in both arrays. In each graph, the solid line represents the fitting result of lattice of 1mm (diameter) obstacles, and the dot dashed line represents the fitting result of lattice of 2mm obstacles. Diamond shaped dots are the experiment result in lattice of 1mm (diameter) obstacles and the square shaped dots are the experiment result in lattice of 2mm (diameter) obstacles.

Finally, we normalize the particle size and the critical offset by the radius of the obstacle, and plot the results in Figure 8. Interestingly, we observed that not only the dependence is the critical offset is linear as the size of the particles but also results for both arrays have similar slope. Although it needs to be validated at the micro scale, this simple correlation would provide the necessary information to tailor the design of DLD systems to specific applications.

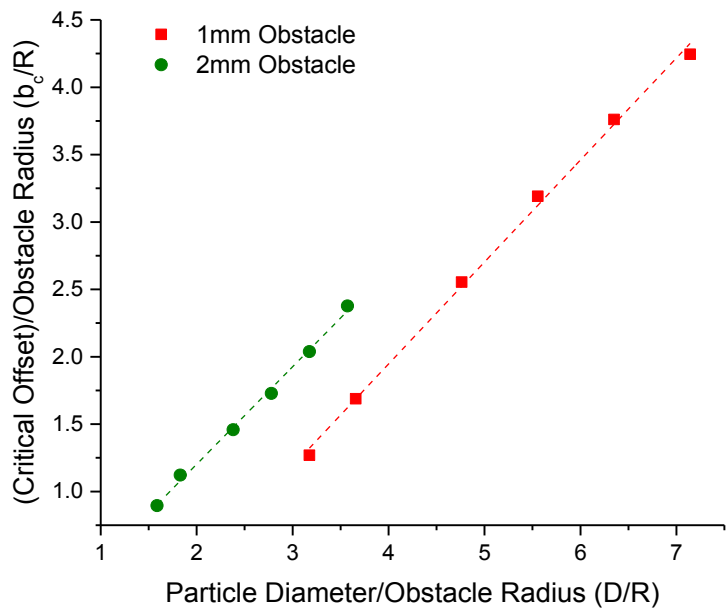

Figure 8 Normalize particle diameter and critical offset $b_c$ both by obstacle radius $R$. For array with 1mm obstacle size, the linear fitting result is $b_c/\mathrm{R} = 0.75696(D/R) - 1.0815, R^2 = 0.99763$. For array with 2mm obstacle the linear fitting result is $b_c/R = 072341(D/R) - 0.24443, R^2 = 0.99828$. The dashed straight lines are linear fitting result for $b_c$ in both arrays.

## Conclusions

For the first time, we use a macromodel to investigate flow-driven DLD microfluidic systems. The use of macromodels allowed us to perform a detailed study of the transport of suspended particles of different size for a wide range of orientations of the average flow with respect to the array of obstacles (forcing angle). We were able to demonstrate the existence of a locked mode for all the particles and different array dimensions. In this locked mode, corresponding to small forcing angles, the migration angle of the particles remains at β=0° until a sudden transition into *zigzag mode* occurs when the forcing angle reaches the critical transition angle. The fact that the transition occurs at increasing angles for larger particles enables particle separation at specific forcing angles. In fact, we observed a linear trend for critical angles as a function of particle size. We note that this observation of a linear trend disagrees with the model originally proposed by Inglis and coworkers [32], which predicts a non-linear dependence (see Figure 4). In addition, we showed that a simple collision model, based on irreversible particle-obstacle interactions, not only captures the sharp locked-to-*zigzag mode* transition but also predicts the migration angles at larger forcing angles. Unfortunately, the prevalent DLD experiments in microfluidics have been focused on small and fixed orientation of the driving flow filed and no general results are available for the behavior of a given size of particles as the forcing angle increases. Therefore, further microfluidic experiments are needed to validate the linear trend observed in the macromodels used here. A possible approach would be to fabricate an array that changes its orientation relative to a main channel [42]. Validating a linear correlation between the size of the species and the critical angle at which they would start to move in *zigzag mode* would enormously simplify the design of flow-driven DLD systems.

**FIGURE 1**

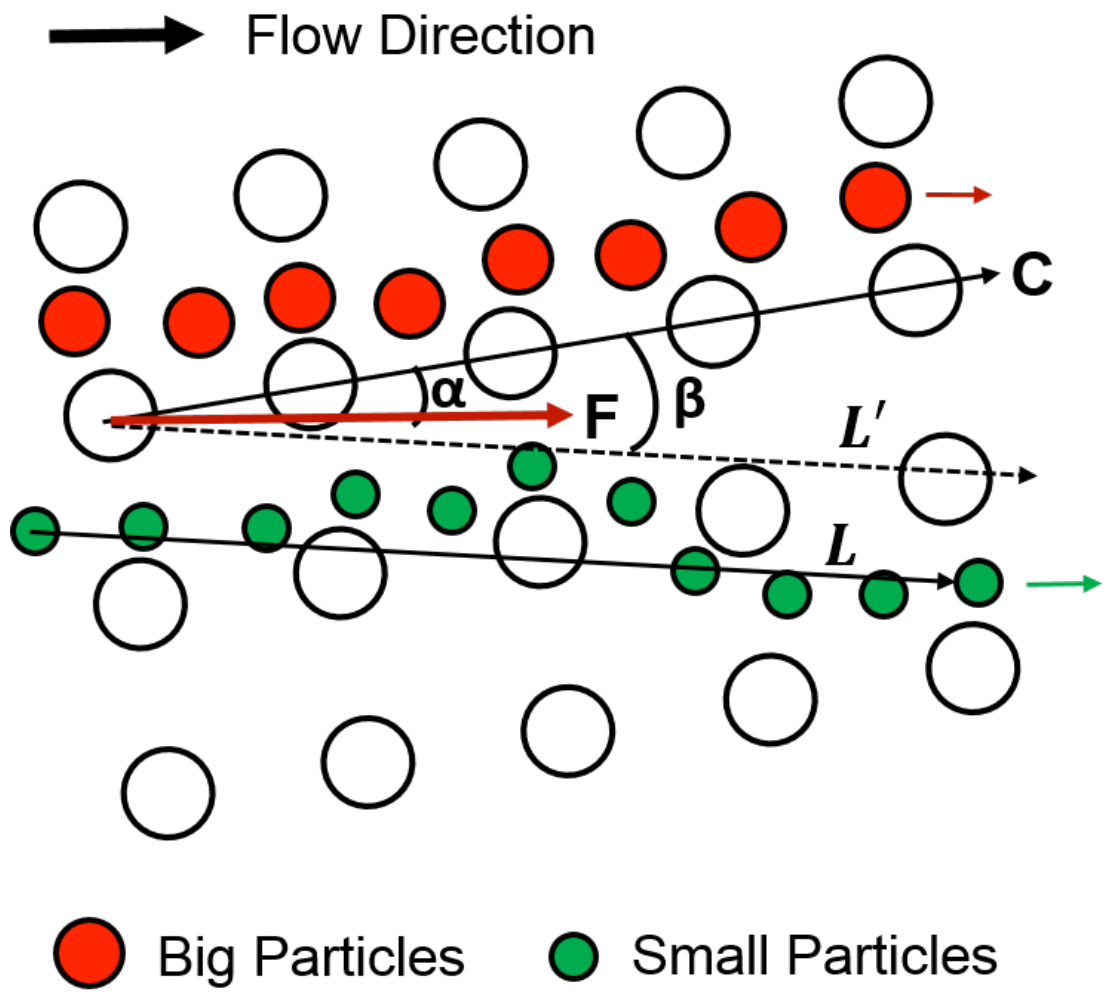

**FIGURE 2**

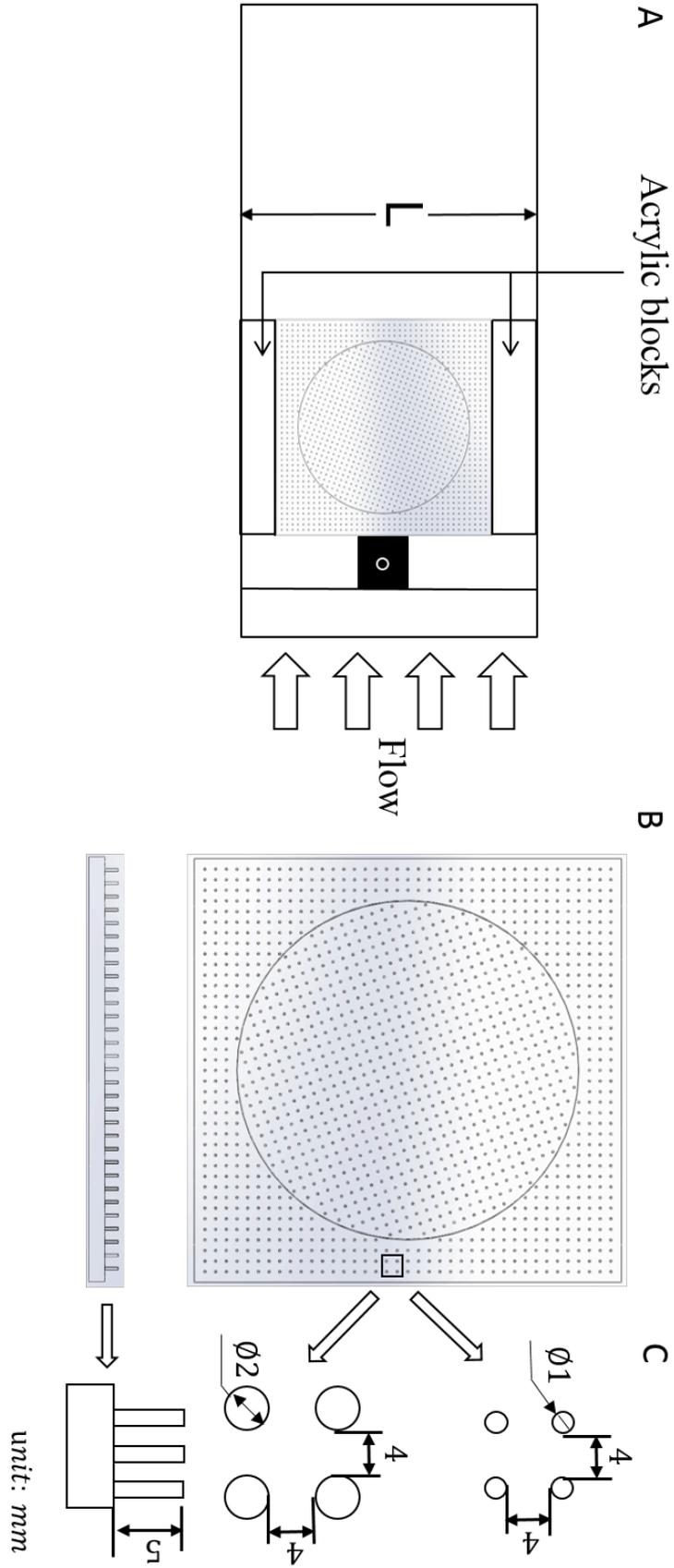

**FIGURE 3**

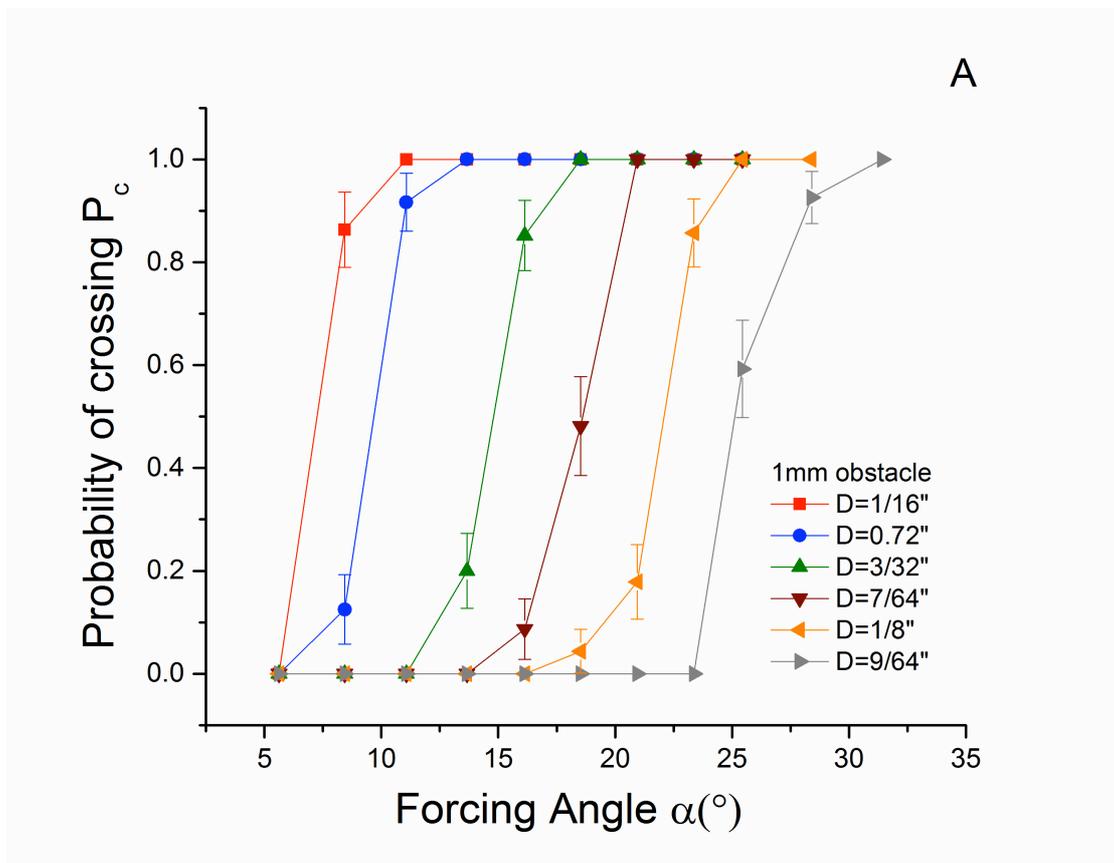

**FIGURE 3**

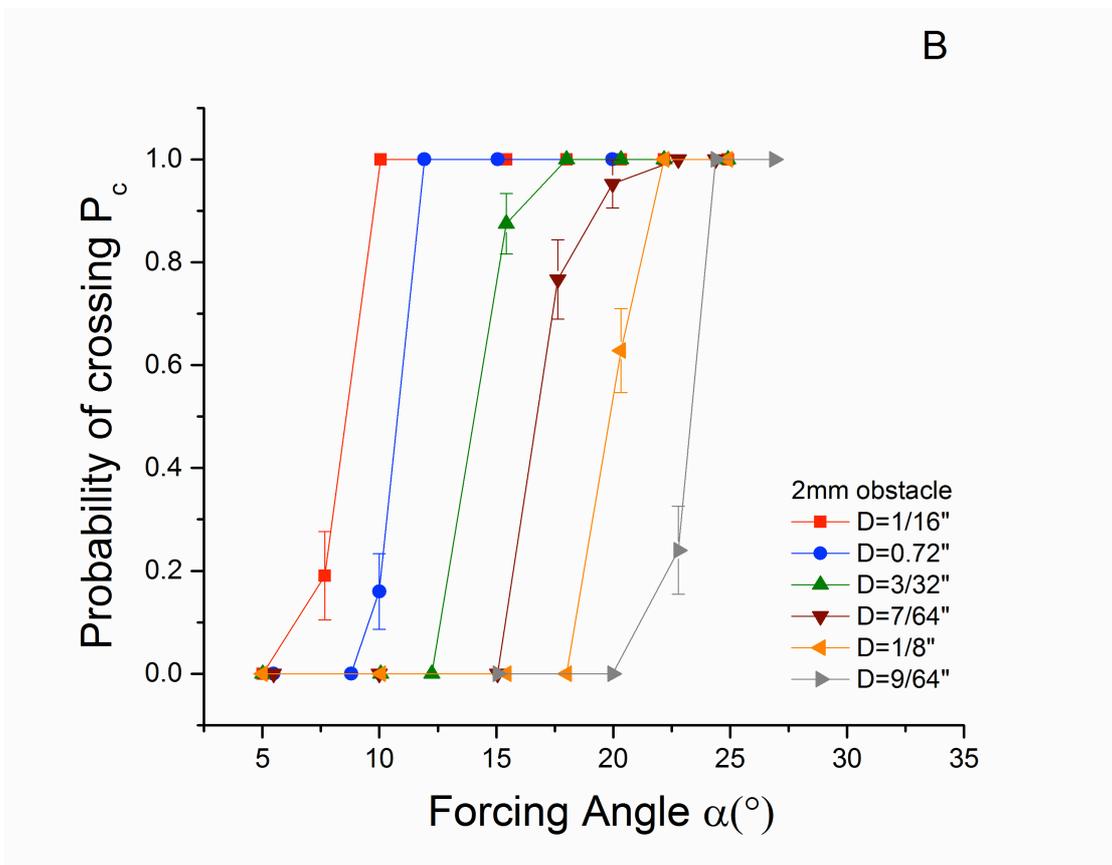

**FIGURE 4**

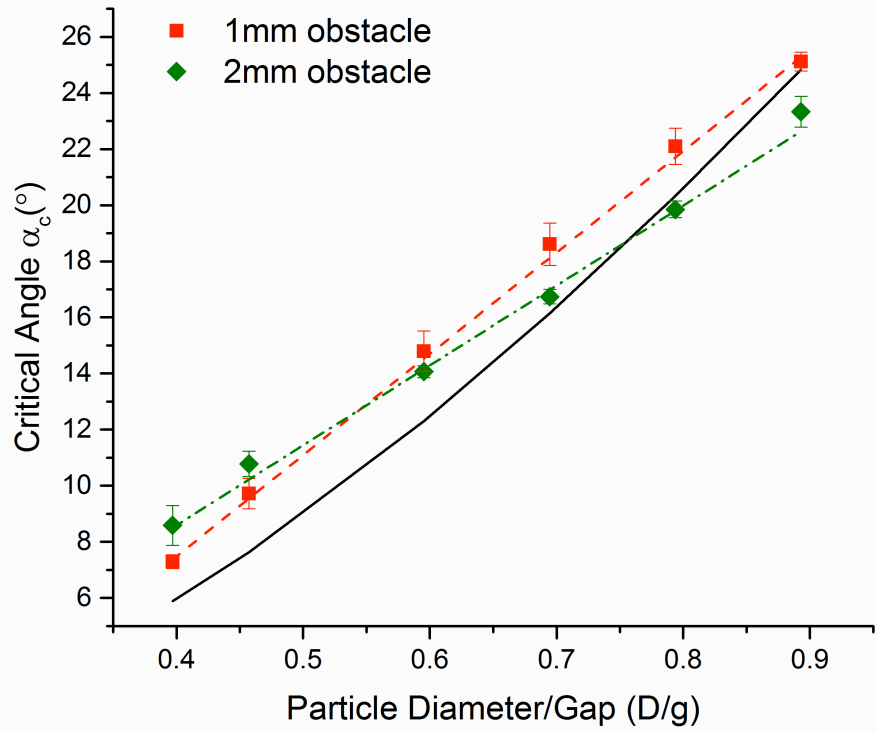

**FIGURE 5**

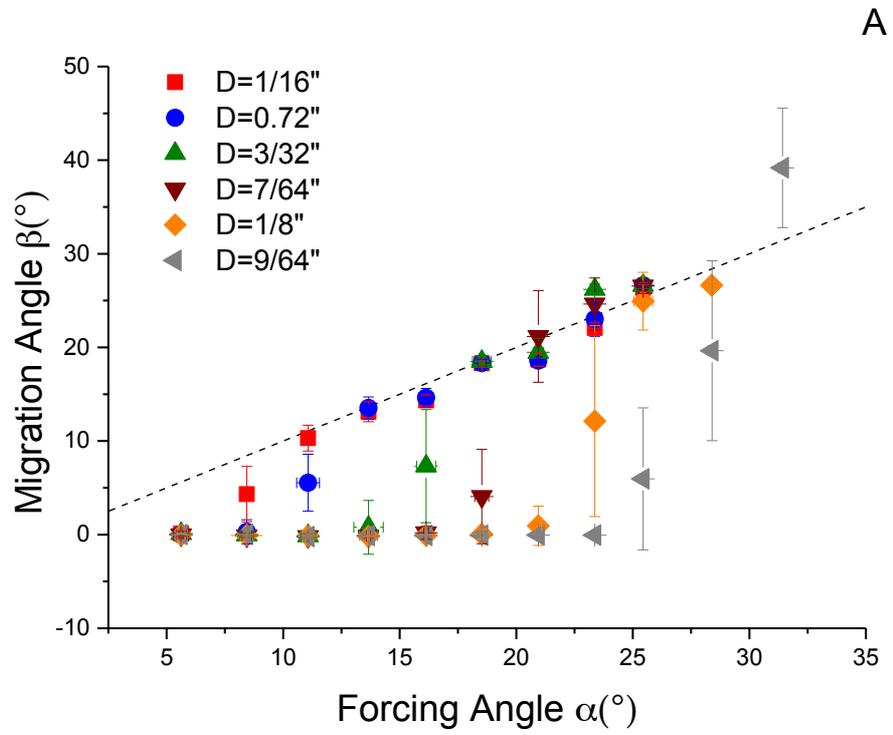

**FIGURE 5**

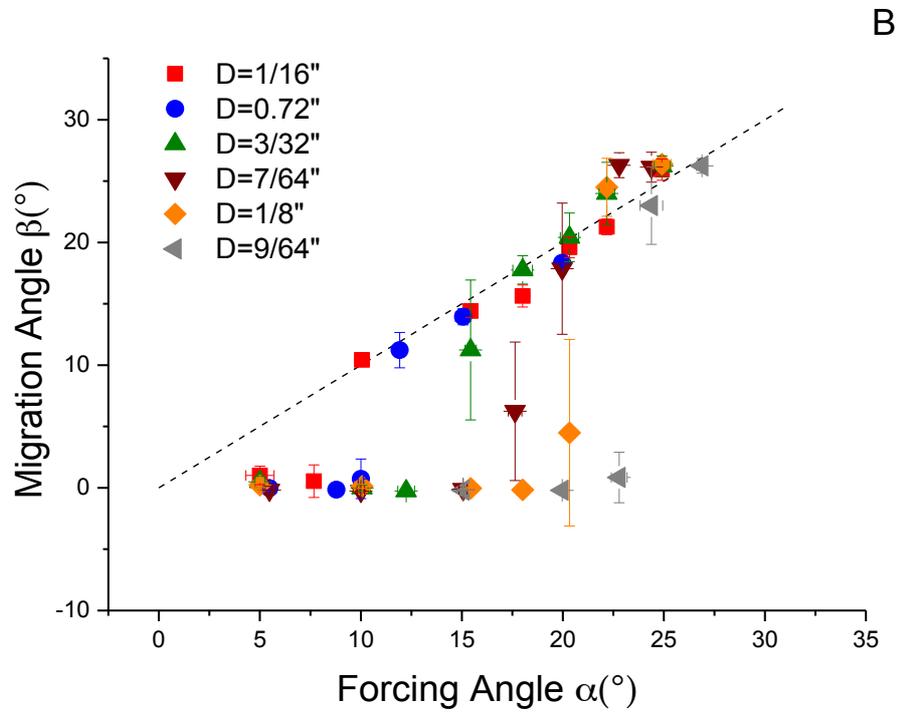

**FIGURE 6**

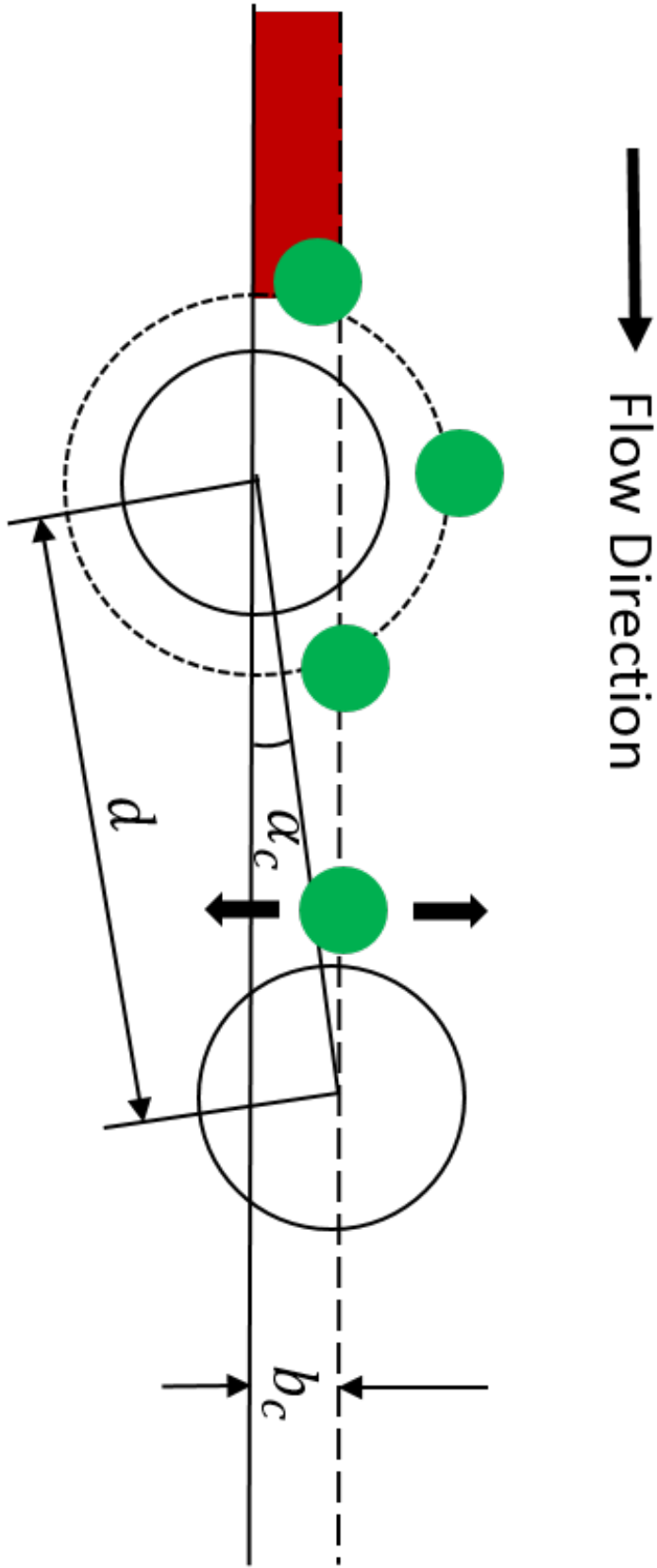

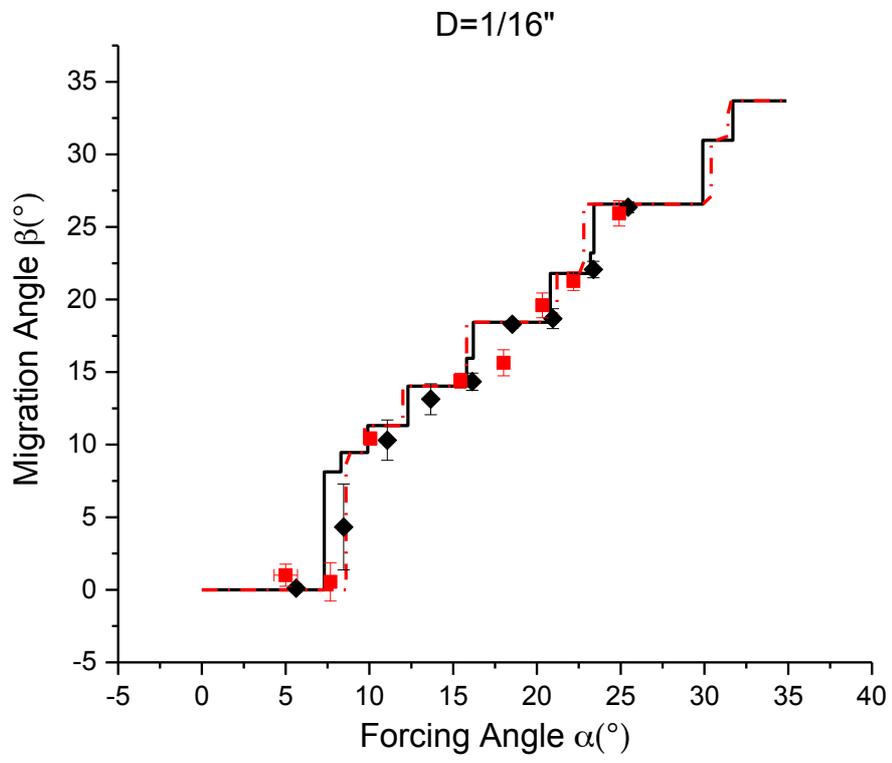

**FIGURE 7**

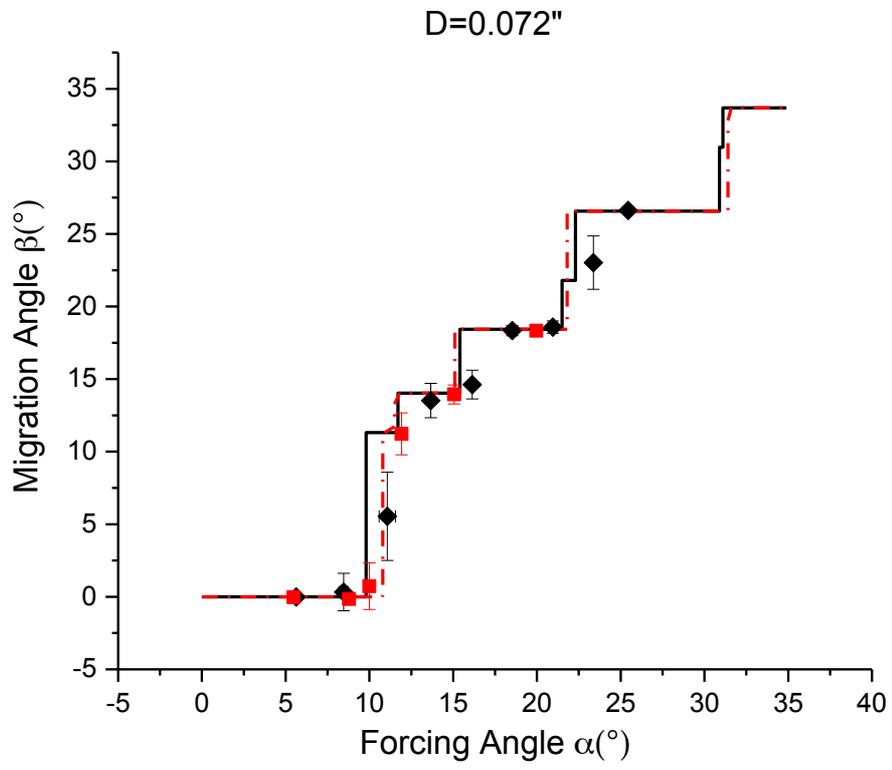

D=0.072"

**FIGURE 7**

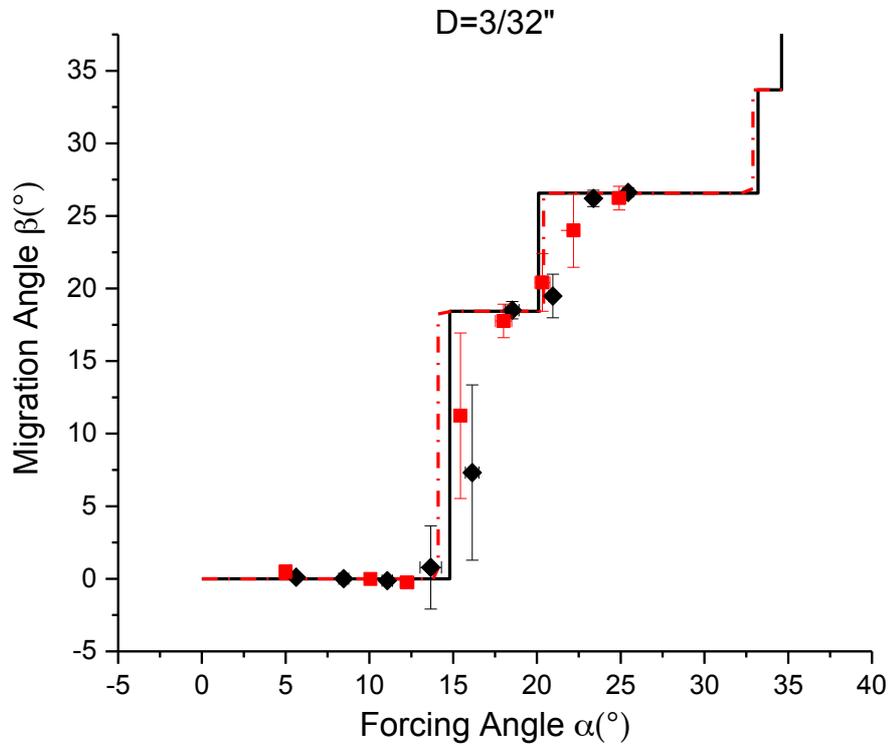

**FIGURE 7**

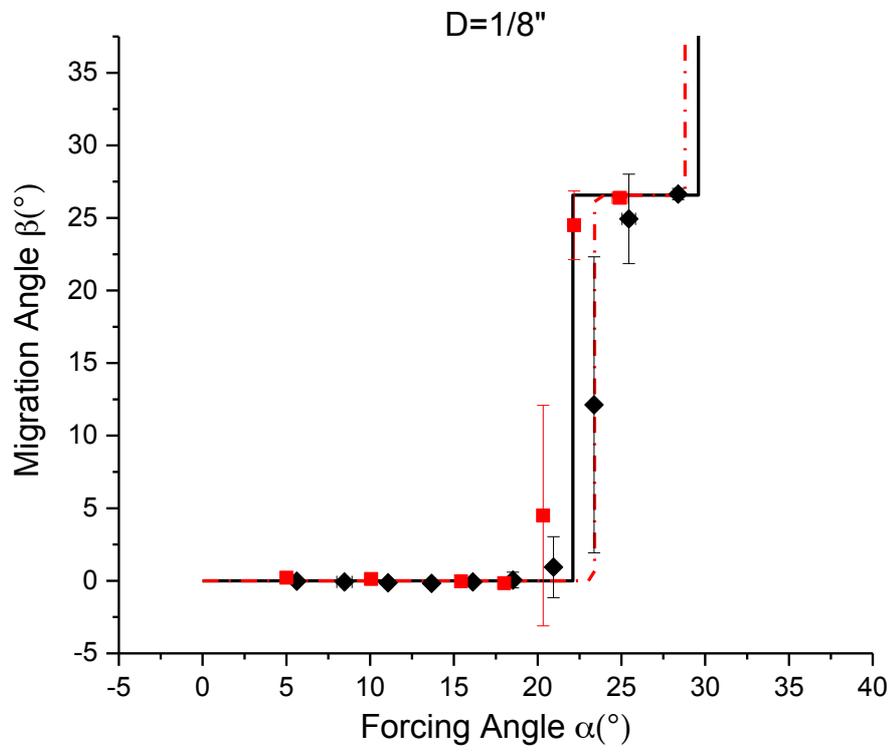

D=1/8"

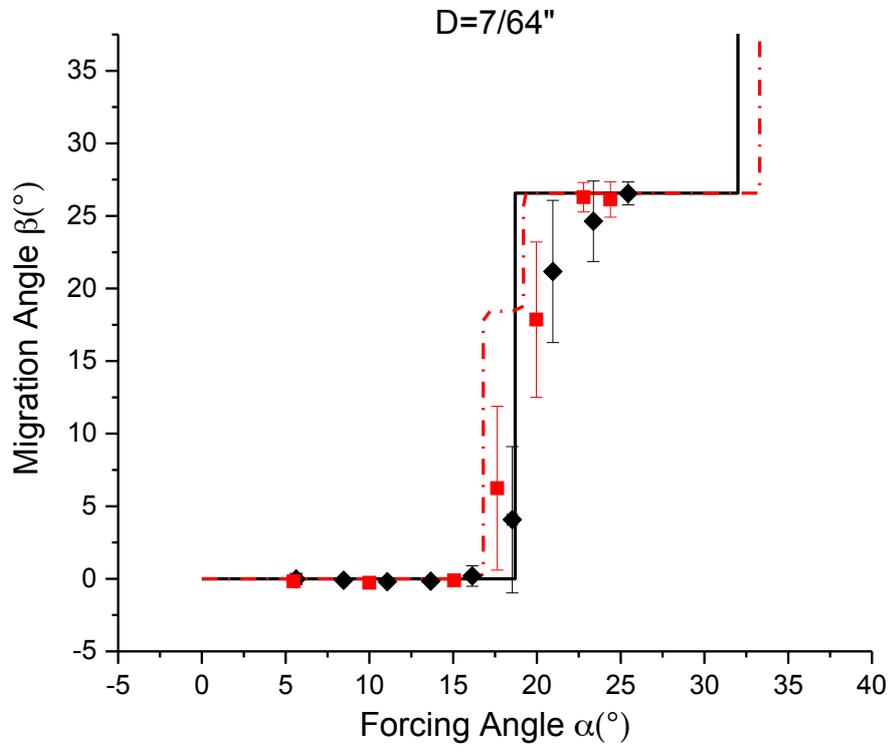

**FIGURE 7**

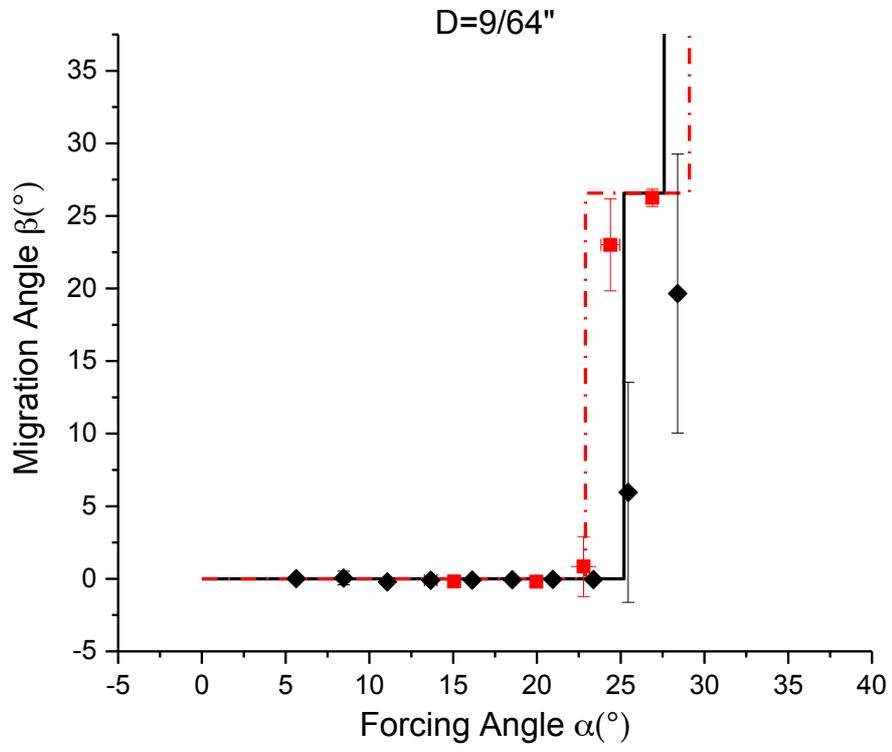

D=9/64"

**FIGURE 8**

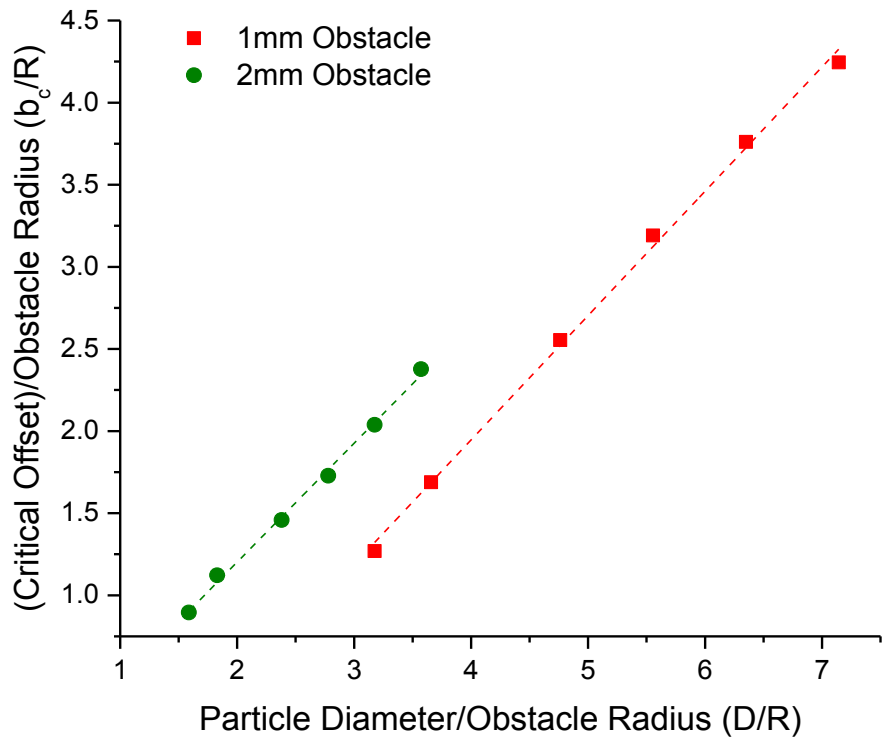

Figure 1. Schematic view of DLD. The circles represent the positions of one particle at different time of movements. The solid line L denotes the direction of particle migration, the arrow F represents the direction of the flow and the solid line that connects centers of a column of obstacles gives the direction of the lattice. The dashed line L' that is parallel to L is drawn to better illustrate the migration angle β.

Figure 2 A. Schematic view of the experiment setup. B. Top and side view of the array of obstacles. C. We use two different arrays with different obstacles diameter as indicated. The height of the obstacles and the surface to surface gap between two obstacles are the same in both arrays.

Figure 3 A. Probability of crossing for different size of particles for 1 mm obstacle array. B. Probability of crossing for different size of particles for 2 mm obstacle array.

Figure 4 Critical angles for different size of particles in both arrays. Particle diameter is normalized by the gap size. Straight lines correspond to a linear fit of the results. The solid curve is calculated using Inglis's model. For 1mm obstacle array, the fitting result is $\alpha_c = 36.2861(D/g) - 6.90216, R^2 = 0.9986$, and for 2mm obstacle array the fitting result is $\alpha_c = 28.7769(D/g) - 2.81352, R^2 = 0.9956$.

Figure 5 A. Migration angle as a function of forcing angle in 1 mm obstacle array. B. Migration angle as a function of forcing angle in 2 mm obstacle array. The dashed line in A and B is $\beta = \alpha$.

Figure 6 Schematic view of a particle obstacle collision showing the case in which the forcing angle is exactly at the critical value. The dashed line depicts the height of the center of a particle coming out of the collision with respect to the centerline of the obstacle, which is exactly the critical offset for this particular particle.

Figure 7 Model fitting results for particles of all sizes in both arrays. In each graph, the solid line represents the fitting result of lattice of 1mm (diameter) obstacles, and the dot dashed line represents the fitting result of lattice of 2mm obstacles. Diamond shaped dots are the experiment result in lattice of 1mm (diameter) obstacles and the square shaped dots are the experiment result in lattice of 2mm (diameter) obstacles.

Figure 8 Normalize particle diameter and critical offset $b_c$ both by obstacle radius $R$. For array with 1mm obstacle size, the linear fitting result is $b_c/R = 0.75696(D/R) - 1.0815, R^2 = 0.99763$. For array with 2mm obstacle the linear fitting result is $b_c/R = 072341(D/R) - 0.24443, R^2 = 0.99828$. The dashed straight lines are linear fitting result for $b_c$ in both arrays.